\documentclass[aps,pra,reprint,superscriptaddress,floatfix]{revtex4-2}

\usepackage[colorlinks=true,urlcolor=blue]{hyperref}
\usepackage{amsmath,amssymb,bm}
\usepackage{graphicx}
\usepackage{float}
\usepackage{cancel}
\usepackage{physics}

\usepackage{tcolorbox}

\begin{document}

\title{Supersolid formation time shortcut and excitation reduction\\
by manipulating the dynamical instability}

\author{Aitor Ala\~na}
\affiliation{Department of Physics, University of the Basque Country UPV/EHU, 48080 Bilbao, Spain}
\affiliation{EHU Quantum Center, University of the Basque Country UPV/EHU, 48940 Leioa, Biscay, Spain}

\date{\today}

\begin{abstract}
    Supersolids are a phase of matter exhibiting both superfluidity and a periodic density modulation typical of crystals. When formed via quantum phase transition from a superfluid, they require a formation time before their density pattern develops. Along this paper some protocols/schemes are proposed for experimental applications, building on earlier descriptions of the role roton instability plays in the supersolid formation process and the associated formation time. {In particular, the Bang-Bang method is put forward as a way to shorten the formation time, and the Parachute Jump scheme is proposed to lessen the excitation produced when crossing the phase transition.} As a case study of the impact that mechanical fluctuations (noise) can have on the phase transition when conducting an experiment, the impact of a mechanical kick before the transition is also investigated. {The proposed schemes achieve a shortening of the formation process for comparable levels of excitation, in the framework of extended Gross-Pitaevskii theory.}   
\end{abstract}

\maketitle

\section{Introduction}

In contrast to other Bose-Einstein Condensates (BECs) in which particles only interact via short-range interactions characterized by the s-wave scattering length, ultracold dipolar gases also exhibit long-range, anisotropic interactions (see, for instance, Refs. \cite{lahaye2009,baranov2012}) thanks to their huge electric or magnetic momenta. A consequence of the dipole-dipole interaction is the so-called roton mode, a local minimum in the dispersion relation at nonzero momentum \cite{Sa03,odell2003,ronen2007,Ch18,Pe19,Sc21}. Depending on the interplay between the density of the gas, its geometry, and the strength of the interactions the roton mode may turn unstable \cite{giovanazzi2002,Ro19} and ignite behaviors such as droplet formation \cite{kadau2016}, supersolids \cite{Ta19,Bo19,Ch19} and other phenomena \cite{wenzel2017,klaus2022}.

The current paper focuses on supersolids, which combine both a superfluid nature, linked with the phase coherence absent in the droplet crystals \cite{Gross62,pitaevskii2016}, and a spatial modulation breaking the translational symmetry typical of cristaline structures \cite{Gross57,cristSov70,boni12,yukalov2020}. Since the first proposals of supersolidity in the past century for solid helium \cite{Le70,chester70} did not achieve a conclusive answer \cite{meisel1992supersolid}, the community has recently drawn its attention towards dipolar gases as a framework to study supersolidity \cite{Zh19,schuster20,blakie2020variational,Ro20,gallemi2020,Bl20,Te21,Zh21,He21b,He21c,alana2022crossing,roccuzzo2022,ilg2023,smith2023}, which has been experimentally proved successful \cite{pollet2019,donner2019,Ta19b,Gu19,Na19,Ta21,He21,Pe21,No21,bottcher2021,sohmen2021,biagioni2022,Bl22,sanchez2023heating}. In dipolar quantum gases the supersolid phase of matter may be reached both by a classical transition from a non-condensed gas at finite temperature to a supersolid \cite{sohmen2021} and a quantum transition from an unmodulated superfluid to a supersolid, whose character has been thoroughly researched both in theoretical and experimental works \cite{Po94,Ma13,Lu15,Ta19,Bo19,Ta19,Se08,Pe21,Ch18}, finding out that the character of the transition can vary from continuous to discontinuous by tuning the transversal trapping frequency \cite{biagioni2022,alana2022crossing}. 

The aforementioned phase transition is between an ordered phase (supersolid) and a disordered one (superfluid), and as such it exhibits clearly distinct behavior when the transition is crossed towards the superfluid or towards the supersolid. Namely, one can relax the supersolid towards the superfluid almost adiabatically with ease, while on the contrary, the supersolid requires a minimum formation time to develop its characteristic density pattern \cite{Bo19,biagioni2022,alana2022crossing,He21}. 

This paper presents proposals to use the special characteristics of the supersolid formation process to our advantage. In an experiment we might desire to speed up the formation process and/or minimize the produced excitation. This may be achieved by altering the evolution of the roton population with a time-dependent scattering length. In particular, accelerating the formation time has a special interest, as it may assist in reducing the effect of three-body losses the condensate may suffer along the transition, effectively extending its useful lifetime.

Although these optimization targets could also be approached within the framework of Optimal Control Theory (see for instance \cite{krotov1993global,sonneborn1964bang}), this work does not aim to provide a general characterization of optimization techniques, rather focusing our efforts on providing experimentally useful and theoretically backed procedures. The shortcut approach to control even though not always optimal has proven effective and we use its general point of view \cite{schaff2011shortcuts,guery2019shortcuts}.

{We propose the Bang-Bang method to reduce the total formation time of the supersolid as well as the Parachute Jump with which the phase transition can be crossed with small excitation without requiring infeasible formation times. Another proposal} arising from this perspective, namely the Kick-Bang, can provide us insight into mechanical noise. Furthermore, it constitutes a proof of principle of how the transition may be affected by perturbation arising from sources such as small vibrations of the traps due to noise in the experimental setup or impacts the condensate may suffer due to non-condensed particles. 

The paper is organized as follows. In Section \ref{sec:system}, an overview of the system under consideration is provided alongside a brief summary of the relevant formulas defining the extended Gross-Pitaevskii theory for dipolar condensates. {In Sec. \ref{sec:BangBang} we present a short review of roton instability emphasizing those aspects relevant to our study \cite{blakie2020variational} and propose the Bang-Bang method. Sec. \ref{sec:Parachute} presents the Parachute Jump proposal.} { In Sec. \ref{sec:KickBang} 
one will find how kicking the condensate before the crossing may also reduce the total formation time.} {In Sec. \ref{sec:3B} the effect of three body losses in the proposed schemes is discussed.} To recap, in Sec. \ref{sec:conclusions} the main results are summarized. 

\section{System}
\label{sec:system}

{Throughout this manuscript a single system has been chosen to be considered, for the sake of simplicity, and despite the fact that the proposals are not specific for such a system. We study a dipolar Bose-Einstein condensate without thermal fluctuations (T=0) trapped in a harmonic trap along the three spatial directions. In particular, we chose an experimental setup previously used by the Dy-lab in Pisa \cite{biagioni2022}.} 

{The effective dimensionality of the condensate is directly related to the trap potential, featuring quasi-1D and quasi-2D scenarios (see Refs. \cite{biagioni2022,alana2022crossing,alana2023formation}). However, dimensionality does not play an important role in the proposals put forward in this manuscript, and thus only the quasi-1D geometry is considered for the sake of briefness.}


{The system is made up of $N=3\times10^{4}$ magnetic atoms of $^{162}$Dy, which have a tunable s-wave scattering length $a_{s}$ and dipolar length $a_{dd}=130 \ a_0$ (where $a_0$ is the Bohr radius). These atoms are trapped by a harmonic potential with frequencies of $(\omega_{x},\omega_{y},\omega_{z})=2\pi\times(15, 101, 94)$ Hz. It is noteworthy that the parameters under consideration were selected based on their experimental feasibility for a specific case. However, it is imperative to bear in mind that the analysis presented herein is conceptually generic and extendable to comparable situations.}


This system can be described in terms of an extended Gross-Pitaevskii (GP) theory including dipolar interactions \cite{ronen2006} and the Lee-Huang-Yang (LHY) correction accounting for quantum fluctuations, within the local density approximation \cite{Li12,wachtler2016,schmitt2016}. The energy functional can be written as $E = E_{\text{GP}} + E_{\text{dd}} + E_{\text{LHY}}$ with
\begin{align}
E_{\text{GP}} &= 
\int \left[\frac{\hbar^2}{2m}|\nabla \psi(\bm{r})|^2 + V(\bm{r})n(\bm{r})+\frac{g}{2} n^2(\bm{r})
\right]d\bm{r}\,,
\nonumber\\
E_{\text{dd}} &=\frac{C_{\text{dd}}}{2}\iint n(\bm{r})V_{\text{dd}}(\bm{r}-\bm{r}')n(\bm{r}') d\bm{r}d\bm{r}'\,,
\label{eq:GPenergy}\\
E_{\text{LHY}} &=\frac{2}{5}\gamma_{\text{LHY}}\int n^{5/2}(\bm{r})d\bm{r}\,,
\nonumber
\end{align}
where $E_{\text{GP}}=E_{\text{k}}+E_{\text{ho}}+E_{\text{\text{int}}}$ is the standard GP energy functional including the kinetic, potential, and contact interaction terms, $V(\bm{r})=(m/2)\sum_{\alpha=x,y,z}\omega_{\alpha}^{2}r_{\alpha}^{2}$ is the harmonic trapping potential, {$n(\bm{r})=N|\psi(\bm{r})|^2$ represents the condensate density (while $\psi$ is normalized to unity)}, $g=4\pi\hbar^2 a_{s}/m$ is the contact interaction strength, $V_{\text{dd}}(\bm{r})= (1-3\cos^{2}\theta)/(4\pi r^{3})$ the inter-particle dipole-dipole potential, $C_{\text{dd}}\equiv\mu_{0}\mu^2$ its strength, $\mu$ the modulus of the dipole moment $\bm{\mu}$, $\bm{r}$ the distance between the dipoles, and $\theta$ the angle between the vector $\bm{r}$ and the dipole axis, $\cos\theta=\bm{\mu}\cdot\bm{r}/(\mu r)$. 
As in Refs. \cite{biagioni2022,alana2022crossing,alana2023formation}, the magnetic dipoles are considered to be aligned along the $z$ direction by a magnetic field $\bm{B}$.
The LHY coefficient is $\gamma_{\text{LHY}}={128\sqrt{\pi}}{\hbar^{2}a_s^{5/2}}/(3m)\left(1 + 3\epsilon_{\text{dd}}^{2}/2\right)$, with $\epsilon_{\text{dd}}=\mu_0 \mu^2 N/(3g)$.

\begin{figure}[t]
 \centerline{\includegraphics[width=0.95\columnwidth]{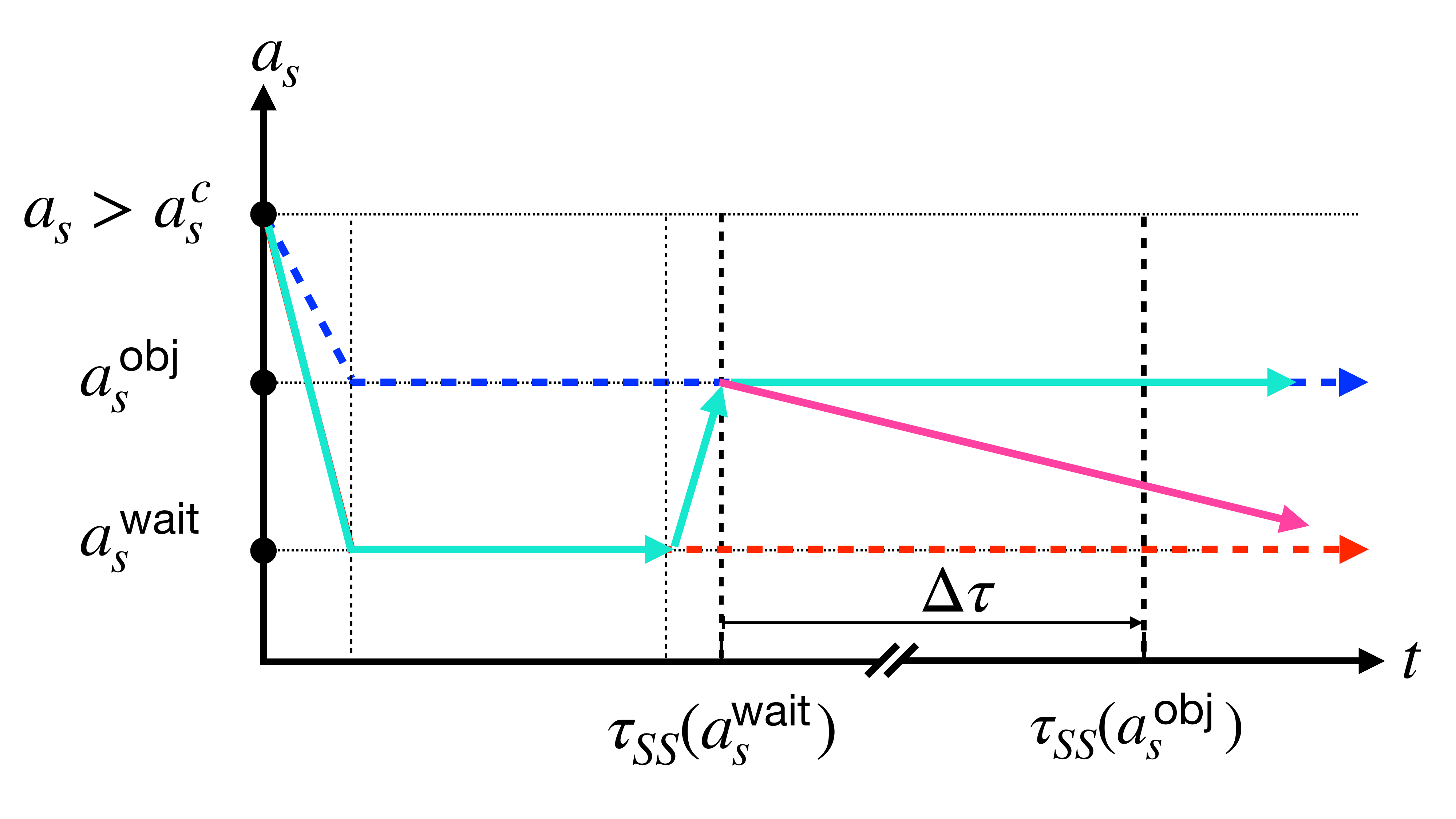}}
 \caption{Scheme of the presented procedures alongside single quenches for comparison. 
 In solid cyan the Bang-Bang, which goes to $a_s^{\text{wait}}$ until almost the formation time of the supersolid $\tau_{SS}(a_s^{\text{wait}})$ to then get back to $a_s^{\text{obj}}$, $\Delta\tau$ indicates how much the supersolid formation is shortened. In solid magenta the Parachute Jump, after forming the supersolid by Bang-Bang the scattering length is linearly reduced back toward $a_s^{\text{wait}}$. Note that the end of the ramp for the Parachute Jump scheme does not necessarily coincide with $\tau_{SS}(a_s^{\text{obj}})$.
 In dashed blue and red a quench directly from the superfluid to $a_s^{\text{obj}}$ and $a_s^{\text{wait}}$ respectively. }
 \label{fig:Scheme}
\end{figure}

{The manuscript considers a phase transition between two states of matter, the superfluid (not modulated) and the supersolid (modulated), which have been proven to be able to exist in the system just described \cite{biagioni2022,alana2022crossing,alana2023formation}. Tuning $a_s$ enables one to perform a second-order transition from one phase to the other, finding the transition point at $a_{s}^{c} \simeq 94.4 a_{0}$ \cite{biagioni2022,alana2022crossing,alana2023formation}. Those are not the only phases available within the parameter space available to the system, for instance, at lower $a_s$ values one would encounter the droplet crystal regime where the fase coherence between droplets is lost, out of the scope of this work.}

We present three different schemes, named Bang-Bang, Parachute Jump, and Kick-Bang respectively. In order to avoid unnecessary complications and facilitate comparisons, in all three schemes we shall start at the same point above the transition, $a_s=a_s^c+1.5 \ a_0$. {The Bang-Bang is presented with the main objective of shortening the total formation time, which is relevant mainly close to the transition point, where the formation is slower. Thus, a convenient choice is to set a supersolid with $a_s^{\text{obj}}$ close to $a_s^c$ as the end goal for the Bang-Bang. The end goal for the Kick-Bang is chosen to be the same $a_s^{\text{obj}}$ to avoid unnecessary complications.}

The Parachute Jump scheme seeks to reduce excitation, which is more relevant once the supersolid gets far from the transition point where the excitation generated by the single-quench is greater. Thus the end goal of such an approach is to finalize the process with a supersolid at a scattering length $a_s^{\text{wait}}$ smaller than $a_s^c$. Along the paper $a_s^{\text{wait}}$ is set to be both the targeted scattering length for the Parachute Jump scheme and the waiting scattering length value of the Bang-Bang scheme (see Sec. \ref{sec:BangBang}), just to avoid introducing unnecessary new parameters. Note however that they are completely independent.

Bang-Bang and Parachute Jump schemes are fairly general and likely to be usefully implemented in the short term, while the Kick-Bang scheme is presented here mainly as a proof of principle and will require further elaboration. The Kick-Bang is also put forward here to highlight the need for a more in-depth investigation of the role of mechanical noise in the supersolid formation time. 

\section{Bang-Bang approach to accelerate the supersolid formation}\label{sec:BangBang}

When the scattering length of a superfluid is reduced below the critical value $a_s^c$ the system starts to dynamically mute into a supersolid, nevertheless, it requires a minimum time before the density turns modulated and breaks the translational symmetry \cite{Bo19,alana2022crossing,alana2023formation}, we refer to the time between the crossing of the transition point and the appearance of a stable modulated density pattern as ``formation time of the supersolid'' {or $\tau_{\text{SS}}$.} 
The current understanding of the supersolid formation process in the superfluid-supersolid transition indicates that it is a consequence of the roton instability. The excitation spectrum of a dipolar Bose gas in a quasi-1D geometry \cite{blakie2020variational},
\begin{equation}
\label{eq:dispersion}
 \hbar\omega_k=\pm\sqrt{\frac{\hbar^2 k^2}{2m}\left(\frac{\hbar^2 k^2}{2m}+2n_0\tilde{V}(k) + 3\gamma_{\text{LHY}}^{(1D)}n_{0}^{3/2}\right)},
\end{equation}
where $\tilde{V}(k)$ denotes the total inter-particle interaction potential in momentum space, exhibits (for some ranges of scattering length) a new local minimum around a finite momentum named roton momentum, $k_{\text{rot}}$ (see e.g. \cite{Pe19}). When the scattering length is reduced below a critical value $a_s^c$ the dipole-dipole interaction overcomes the repulsive contact interaction and Eq.\eqref{eq:dispersion} becomes imaginary for some momenta around $k_{\text{rot}}$, i. e., some momentum states become unstable and experience a population growth. 
The supersolid is formed when the roton momentum state is macroscopically occupied, breaking the translational symmetry of the condensate by a close to sinusoidal pattern, whose wavelength is associated with the roton mode, $\lambda=2\pi/k_{\text{rot}}$ (see e.g. \cite{Ta19,Ta19b,Ta21,Ch18,Ch19}). 

Both experimental and theoretical works \cite{Ch18,Ch19,blakie2020variational,alana2023formation} have shown that the population increase of the roton mode is exponential in the initial stages of the phase transition, slowing down once the modulation is formed and quantum fluctuations stabilize the system. 
To study the speed of the formation one can ignore other unstable momenta and only consider the roton momentum, related to the highest value of $|\text{Im}[\omega_k]|$, which corresponds to the momentum state whose population grows the fastest. The imaginary part of the roton frequency defines a timescale $\tau_R(a_{s})$ decreasing with lowering scattering lengths, 
which has been found to be proportional to the total formation time $\tau_{\text{SS}}$ \cite{alana2023formation}. 

Now suppose we are interested in obtaining a supersolid with scattering an objective length $a_s^{\text{obj}}$ below, but close to, the critical value $a_s^c$, which would need a long formation time since close to the transition point the imaginary part of $\omega_k$ is small. This may be interesting, for instance, to study supersolids with weak spatial modulation. 

Drawing from current knowledge about the formation process one can propose the following ``shortcut'' to the supersolid formation: The superfluid is quenched to a scattering length $a_s^{\text{wait}}$ smaller than the target $a_s^{\text{obj}}$, and kept ``waiting'' at that value, $a_s^{\text{wait}}$, almost until the formation time required with such scattering length, $t_{{\text{wait}}}\lesssim \tau_{\text{SS}}(a_s^{\text{wait}})$. After an evolution of $t_{{\text{wait}}}$ the scattering length is quenched again to the target value (see the cyan line in Fig. \ref{fig:Scheme}).

\begin{figure}[t!]
 \centerline{\includegraphics[width=1.0\linewidth]{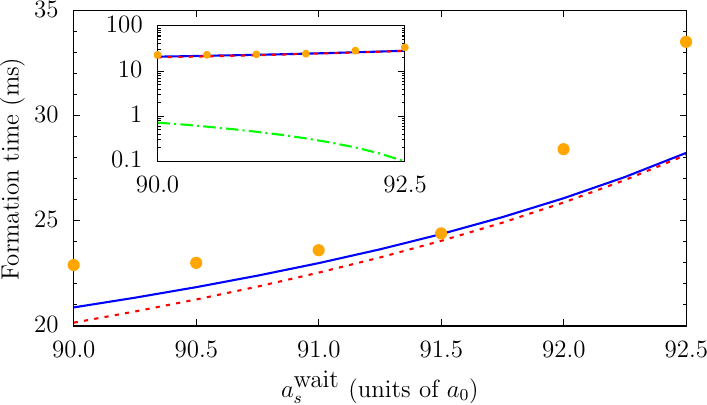}}
 \caption{{$\tau_{\text{SS}}^{\text{BB}}$ (solid blue) and the formation time's term given by the time spent in $a_s^{\text{wait}}$ (dashed red), predicted with the quasi-1D model used in \cite{blakie2020variational,alana2023formation} with $a_s^{\text{obj}}=a_s^c-1.0a_0$ just as for the shown simulations. Points indicate simulation results for the total time of the Bang-Bang, where the second bang was performed a safety  time (of 3 ms) before the formation times $\tau_{\text{SS}}(a_s^{\text{wait}})$. The inset shows, in logarithmic scale, the same information together with the time required in $a_s^{\text{obj}}$ for the formation to be completed according to the second term of Eq. \ref{eq:BB_time} (dotted-dashed green).}}
 \label{fig:BB_time}
\end{figure}

In this scheme, the total formation time needed to form the supersolid in the condensate can be divided between  $t_{\text{wait}}$, the time in which the system is with $a_s^{\text{wait}}$, and $t_{\text{obj}}$, the time in which the system is with $a_s^{\text{obj}}$. The former of both times, $t_{\text{wait}}$, is much larger than the latter, $t_{\text{obj}}$, {if $a_s^{\text{wait}}$ is close to $a_s^{\text{obj}}$ (see Fig. \ref{fig:BB_time})}. At the time in which the scattering length is suddenly changed to the desired value $a_s^{\text{obj}}$, when $\tau=t_{\text{wait}}$, the roton mode is populated enough so as to be immediately noticeable. In the rest of the evolution period, $t_{\text{wait}}<\tau<t_{\text{wait}}+t_{\text{obj}}$, the already noticeable modulation grows until stabilizing around an equilibrium value. This ``shortcut'', called Bang-Bang because it is implemented by two quenches, allows us to create a supersolid with $a_s^{\text{obj}}$, but waiting only close to the duration of the formation time characteristic of $a_s^{\text{wait}}$. 

Although the roton momentum slightly changes at varying the scattering length, if {$|a_s^{\text{obj}}-a_s^{\text{wait}}|$ is around a couple of $a_0$} the shift should be small enough so as to ensure that after the second quench the characteristic roton $k_{\text{rot}}(a_s^{\text{obj}})$ is also sufficiently populated. {The effect of the changing roton momentum and its relation to $a_s^{\text{wait}}$ is discussed at the end of the section by using the results of the quasi-1D infinite model alongside results from 3D simulations.}

\begin{figure}[t]
 \centerline{\includegraphics[width=1.\linewidth]{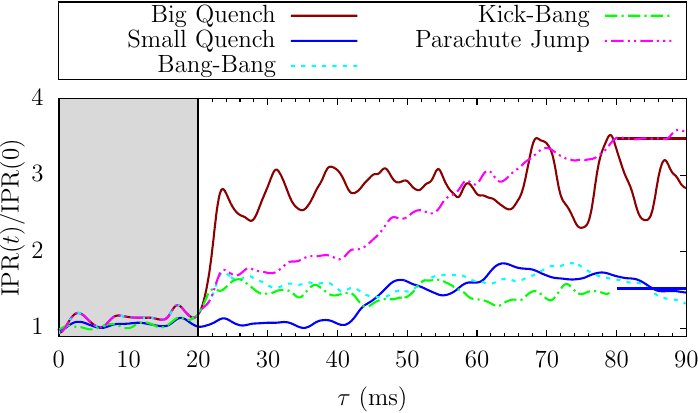}}
 \caption{Inverse participation ratio (with respect to the initial value) along the time evolution{, where $\tau$ is the time elapsed after the first quench (equivalently, the time elapsed in the supersolid regime). The plots follow the color scheme shown in the key box above: Single quench scheme for final scattering value $a_s^c-3.5 \ a_0$ (``Big Quench'') and $a_s^c-1.0 \ a_0$ (``Small Quench'') and Kick-Bang (Sec. \ref{sec:KickBang}), Bang-Bang (see Sec. \ref{sec:BangBang}) and the Parachute Jump (see Sec. \ref{sec:Parachute}) schemes.}
 The 20 ms before the first quench in the case of the Kick-Bang scheme, which is left to evolve after the kick some time before quenching, is not shown. The second quench for the Bang-Bang occurs at $\tau=20$ ms, denoted by a color change in the background (from gray to white). The horizontal lines at the right show the IPR for the ground state configurations at $a_s^c-3.5 \ a_0$ and $a_s^c-1.0 \ a_0$. Note that for $\tau\ge20$ ms Bang-Bang and Kick-Bang indicate supersolids at $a_s^c-1.0 \ a_0$, while the magenta one changes the scattering length gradually between $a_s^c-1.0 \ a_0$ (at 20 ms) and $a_s^c-3.5 \ a_0$ after the initial ``solidification''.}
 \label{fig:IPR}
\end{figure}

\begin{figure}[t!]
 \centerline{\includegraphics[width=1.0\linewidth]{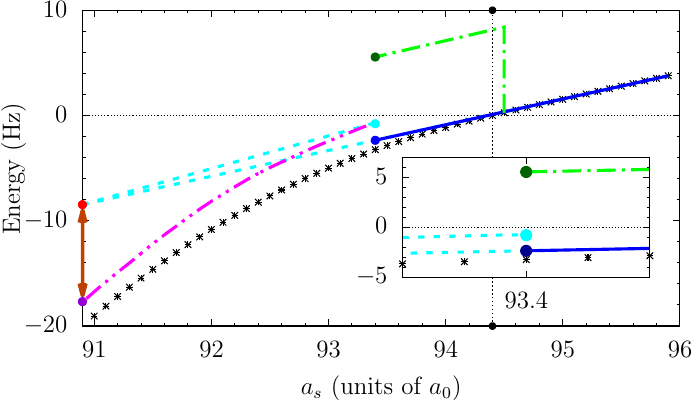}}
 \caption{The behavior of the energy normalized respect to the value at the transition point, $(E[a_s] -E[a_s^c])/h$ (where $E[a_s^c]/h$ corresponds to $432.2$ Hz) for the ground state (black stars), a single quench to $a_s^{\text{obj}}$, Bang-Bang scheme and the Kick-Bang scheme following the color classification of Fig. \ref{fig:Energy}, where points have been introduced as an indication of the endpoints. The inset shows a zoom around the $a_s^{\text{obj}}$ to make the energy difference of the {energies after Kick-Bang, Bang-Bang, and single quench schemes} clear. 
 {Only the last step of the Parachute Jump scheme (the reduction of the scattering length to $a_s^c-3.5 \ a_0$ with a constant speed $da_s/dt=0.04 \ a_0/$ms) is shown, with the color-scheme of Fig.\ref{fig:IPR}. This is done because the previous step is to apply the Bang-Bang, already shown.}
 It is apparent that the Bang-Bang creates an ``excess excitation'' as its endpoint, the cyan point, is higher than the endpoint of a single quench, the blue point. Meanwhile, the transition performed after a kick exhibits a sudden high excitation due to the kick which is not able to dissipate afterward, ending the process with higher excitation (see the location of the dark-green point with respect to the others). Regarding the Parachute Jump there is a significant energy reduction remarked with the double-headed dark-orange arrow between the energy after a single quench (red dot) and after the Parachute Jump scheme (dark-violet dot).}
 \label{fig:Energy}
\end{figure}

After one unstable momentum state starts to be macroscopically occupied the system leaves the linear excitation regime and the computed Bogoliubov excitation spectrum of Eq.\eqref{eq:dispersion} breaks down (the approximation $n(x)\approx n_0$ does not hold anymore) \cite{Ch19,blakie2020variational}. As a consequence, the roton momentum driving the modulation instability is no longer guaranteed to be the characteristic momentum of the system, as neither is, for instance, the wavelength of the ground state; thus, the wavelength of the supersolid may be slightly shifted from the roton one. This is manifested by the newly created maxima moving from their creation positions, associated with the roton, and changing the relative distances (see Sec. \ref{sec:Parachute}). It is also visible in the momentum density distribution after the formation of the supersolid pattern (see e.g. \cite{alana2023formation}). 

Fig. \ref{fig:IPR} shows an example of the effectiveness of the Bang-Bang approach to shorten the formation time. 
{Following \cite{alana2023formation} the inverse participation ratio (IPR) is used as a proxy for crystallization. The IPR is a measure of how localized a system is \cite{calixto2015inverse,cuevas2002critical,murphy2011generalized}, in the context of superfluid-supersolid phase transition how localized it is spatially. Although generalizations exist, a simple definition of IPR suffices for this transition, namely, the integral over all space of probability density ($|\psi|^2$ instead of $N|\psi|^2$) square, $\int |\psi|^4dV$ \footnote{{More generally one can consider a system well described by an orthonormal basis $\{\ket{i}\}_i$ such that any state is given by $\ket{\psi}=\sum_i p(i)\ket{i}$ with $p(i)$ as the probability to inhabit the eigenstate. For such system $\text{IPR}=\sum_ip(i)^{2}$. Thus, if a state only ``participates'' in one state, $\ket{\psi}=\ket{1}$ the IPR is 1 while if it participates in $N$ states with probability $p(i)=1/N$ the IPR is $1/N$ (less localized).}}. }

\begin{figure*}[t!]
    \centerline{\includegraphics[width=2\columnwidth]{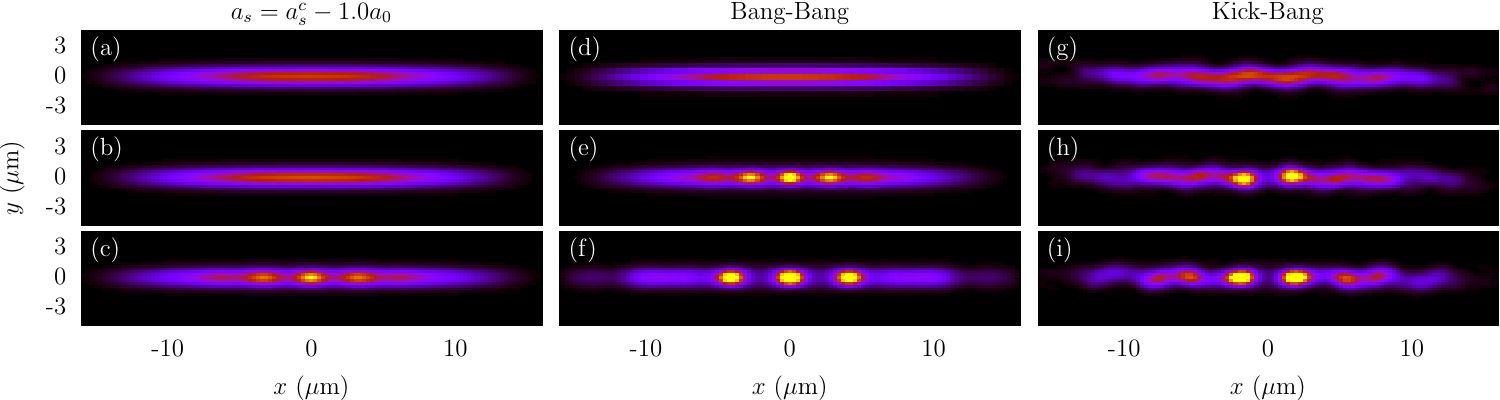}}
\caption{Comparison of the density configurations when applying a single quench (left) and the proposed shortcut schemes (center and right) for $\tau=0$ ms (a, d and g), $\tau=20$ ms (b, e and h), $\tau=40$ ms (c, f and i). Note that for the Kick-Bang $\tau=0$ ms represent $20$ ms after the kick. {The color scheme is the same for all figures as well as Fig.\ref{fig:dens_parachute} below.}}
\label{fig:dens_shortcut}
\end{figure*}

{The validity of the IPR as an indicator of the supersolid structure resides in the significantly different localization levels of superfluids and supersolids. The latter are significantly more localized as the particles pack closely around the separated clusters instead of spreading out. IPR also shows information about dynamical excitations of the system that localize/delocalize it. To extract the supersolid formation time from simulation data we follow the criterion of Ref.\cite{alana2023formation} for which the time until the first local maximum of IPR after the exponential increase is used. Find in \cite{calixto2015inverse} another example of IPR in the context of phase transitions.}

%

Dark-red and blue solid lines in Fig. \ref{fig:IPR} represent the evolution after a single quench to $a_s^c-3.5 \ a_0$ and $a_s^c-1.0 \ a_0$ respectively, while the dashed cyan line is the result of the Bang-Bang scheme with $a_s^{\text{wait}}=a_s^c-3.5 \ a_0$ and $a_s^{\text{obj}}=a_s^c-1.0 \ a_0$. A clear shortening of the formation time is observed, as the total formation time is shorter in the Bang-Bang approach than in the simple quench approach by a factor of around 2 (around 20 ms instead of around 50 ms).

Energywise, as long as no dissipation methods are present, the Bang-Bang scheme is likely to get the system slightly more excited than with a single quench. This happens because the energy change of the system after a quench is \cite{alana2022crossing},
\begin{equation}
    {\Delta E \approx \frac{2\pi\hbar^2}{m}\Delta a_{s} {\cal E}^{\text{int}}[n],}
\end{equation}
where ${\cal E}^{\text{int}}[n]=\int |\psi|^4 dV=\text{IPR}$. Note that only when $a_s$ changes can the system change its energy, thus, the energy is constant between the two quenches. It is straightforward to see that the Bang-Bang scheme has an excitation of, 
\begin{equation}\label{eq:excite}
    {\frac{a_s^{\text{obj}}-a_s^{\text{wait}}}{m/2\pi\hbar^2}\left({\cal E}^{\text{int}}[n(t_{\text{wait}})]-{\cal E}^{\text{int}}[n(0)]\right)},
\end{equation}
additional to the excitation a single quench scheme would produce. The simple infinite quasi-1D models predict only the growth of the roton mode \cite{blakie2020variational,Ch19}, which would give near zero excess excitation as long as the second quench is applied before the macroscopical occupation of $k_{\text{rot}}$. {It is convenient to perform the second quench a ``safety  time'' $\Delta t$ ---close to the time interval between when the exponential growth of the roton population starts to be noticeable in the IPR and the end of such growth, which is less or equal to 3 ms for all the performed simulations--- before the formation time at the waiting scattering length value.  This is convenient to avoid high excitation because otherwise, one quenches the system with already formed [$t_{\text{wait}}=\tau_{\text{SS}}(a_s^{\text{wait}})$] or almost formed supersolid structure. Thus, to avoid the risk of high excitation $t_{\text{wait}}=\tau_{\text{SS}}(a_s^{\text{wait}})-\Delta t$.} 

However, a finite 3D system also exhibits dynamical responses to the quenches other than the roton mode \cite{alana2022crossing}, which can be responsible for a nonzero excess excitation \footnote{${\cal E}^{\text{int}}$ changes after the first quench, depending on its value at the moment of the second quench the excitation will vary. The excess excitation could be negative (the system will be less excited than for the single-quench scheme) if ${\cal E}^{\text{int}}[n(t_{2^{\text{nd}}\text{Bang}})]< {\cal E}^{\text{int}}[n(t_{1^{\text{st}}\text{Bang}})]$}. We depict in Fig. \ref{fig:Energy} the behavior of the energy {and in Fig. \ref{fig:dens_shortcut} the density distributions in selected times} when this scheme ---and the schemes of Secs. \ref{sec:KickBang}-\ref{sec:Parachute}--- is applied. 

The scheme's efficacy resides in two properties: i) that the growth of the roton mode is faster with smaller scattering length values; ii) that the momentum states populated during the waiting time are close to (overlap significantly with) the unstable momenta at the desired scattering length. Notice that there is no need for the waiting configuration to be inside the supersolid regime; one may, for instance, implement it with $a_s^{\text{wait}}$ that would not support a supersolid, namely, if we lower further the scattering length we would find ourselves in the droplet crystal regime, where the phase coherence and superfluidity are lost. 


{One could be interested in optimizing the formation process within the aforementioned limitations. To do this one must take into account that to each $a_s^{\text{wait}}$ it corresponds a characteristic roton momentum $k_{\text{rot}}(a_s^{\text{wait}})$ given by the maximum instability in Eq.\eqref{eq:dispersion}, which is related to the formation time of the supersolid\cite{alana2023formation}. When the system ``waits'' in $a_s^{\text{wait}}$ not only the roton but also close momenta get unstable and grow exponentially, among which we are interested in the characteristic momentum of the desired supersolid, $k_{\text{rot}}(a_s^{\text{obj}})$.}

{As previously studied (see \cite{alana2023formation}) the supersolid formation time for a given scattering length is $\tau_{\text{SS}}(a_s)=\alpha\tau_{R}(a_s)$ where $\alpha$ is a numerical constant dependent on geometry and initial momentum distribution (estimated in 6.5 for this system) and,}
\begin{equation}
    {\tau_{R}(a_s)=\frac{1}{\text{Max}_{k}\left[\left|\text{Im}[\omega_k(a_s)]\right|\right]},}
\end{equation}
{is the inverse frequency of the most unstable mode given by Eq.\eqref{eq:dispersion}. Since the evolution of the roton and the other unstable momentum states is exponential, the population growth of each mode will be given by,}
\begin{equation}
    {P_k(\tau_{\text{SS}})/P_k(0)=\text{exp}\left[2{\tau_{\text{SS}}(a_s)/\tau_{k}(a_s)}\right]}
\end{equation}
{where $\tau_{k}(a_s)$ is the inverse of the imaginary part of the frequency $k$ at the given scattering length and the 2 factor arises from the population being the square of the module. Hereinafter momenta corresponding to rotons for $a_s^{\text{obj}}$ and $a_s^{\text{wait}}$ are denoted with $R(\text{obj})$ and $R(\text{wait})$ respectively instead of a general $k$. Therefore, after waiting for $\tau_{\text{SS}}(a_s^{\text{wait}})$ the population growth of the objective roton will have been dictated by $\tau_{R(\text{obj})}(a_s^{\text{wait}})>\tau_{R(\text{obj})}(a_s^{\text{obj}})$.}

{The total formation time of the Bang-Bang can be divided in the supersolid formation time for the waiting scattering length $[t_{\text{wait}}\lesssim\tau_{\text{SS}}(a_s^{\text{wait}})]$ and the time it inhabits in $a_s^{\text{obj}}$ until the formation ends ($t_{\text{obj}}$). Considering the upper bound for $t_{\text{wait}}$ and the exponential growth of the momentum states it follows that,}
\begin{equation}
    {\frac{\alpha\tau_{R(\text{wait})}(a_s^{\text{wait}})}{\tau_{R(\text{obj})}(a_s^{\text{wait}})} + \frac{t_{\text{obj}}}{\tau_{R(\text{obj})}(a_s^{\text{obj}})}=\alpha\tau_{R(\text{obj})}(a_s^{\text{obj}}),}
\end{equation}
{and thus, the lower bound for the total formation time of the Bang-Bang scheme, $\tau_{\text{SS}}^{\text{BB}}$, is given by,}
\begin{align}\label{eq:BB_time}
    {\frac{\tau_{\text{SS}}^{\text{BB}}}{\alpha} }&{=\tau_{R(\text{wait})}(a_s^{\text{wait}})+\frac{\tau_{R(\text{obj})}(a_s^{\text{obj}})}{\tau_{R(\text{obj})}(a_s^{\text{wait}})}}\nonumber\\
    &{\times \left[\tau_{R(\text{obj})}(a_s^{\text{wait}})-\tau_{R(\text{wait})}(a_s^{\text{wait}})\right],}
\end{align}
{which may increase when reducing $t_{\text{wait}}$.}

{Using the quasi-1D model used in \cite{blakie2020variational,alana2023formation} one sees that although the first term of Eq.\eqref{eq:BB_time} decreases when lowering $a_s^{\text{wait}}$, the second one increases\footnote{{The model works better with $a_s^{\text{wait}}$ inside the supersolid regime. It underestimates a bit the formation times when it is close to the transition point, 
and it overestimates the values for very low $a_s$, however providing good agreement a couple of Bohr radii below transition, e.g. $\tau_{\text{SS}}(91.0a_0)=22$ (really close to the simulations). Nevertheless, it correctly predicts the behavior of the system qualitatively, as also shown by \cite{blakie2020variational,alana2023formation}. The divergence may be explained by excitations, since both close and way below the transition get strong respect to the modulation of the supersolid ground state (which close to the transition is weakly modulated). However, given the simplicity of the quasi-1D infinite model, it predicts remarkably well the behavior of the transition.}}. The total formation time decreases at the beginning (see Fig. \ref{fig:BB_time}) to later slow down when the second term becomes more relevant and compensates for the reduction of the first one. }

{The time inhabiting $a_s=a_s^{\text{wait}}$ is limited by $\tau_{\text{SS}}(a_s^{\text{wait}})$, the formation time of the supersolid at that scattering length. This is so because the second quench must be performed before allowing the system to develop the supersolid structure, to avoid unnecessary excitation [see Eq.\eqref{eq:excite}]. When $a_s^{\text{wait}}$ and $a_s^{\text{obj}}$ are not too different both roton modes grow at a similar pace, and thus after $\tau_{\text{SS}}(a_s^{\text{wait}})$ both rotons are populated enough. However, when $a_s^{\text{wait}}\ll a_s^{\text{obj}}$ both rotons develop at different pace, inducing a non-ideal population of $R(a_s^{\text{obj}})$ after the waiting time, which must be compensated by time waiting in $a_s=a_s^{\text{obj}}$.}

{The improvement one gets from reducing $a_s^{\text{wait}}$ diminishes gradually. Thus, it is convenient to keep the values below but relatively close to $a_s^{\text{obj}}$ as done in the examples provided in this paper. Although there is nothing preventing us from decreasing $a_s$ to very low values to get the extra reduction in the formation time, this would generate more excitations which may not be desirable. Indeed, performed simulations show that it gets more and more difficult to apply the scheme for lower $a_s$ values, since the collective oscillations get stronger because of the high excitation [see Eq.\eqref{eq:excite}].}

\section{The Parachute Jump approach to excitation reduction}\label{sec:Parachute}

Shortening the supersolid formation time is a relevant topic for experimental realizations of supersolids and possible applications. This is also the case for the reduction of excitation created when crossing the phase transition, as it may assist in reducing undesired effects and enable the study of subtle phenomena that could be unobservable for excited states. Adiabatically crossing the phase transition from superfluid to supersolid requires really slow scattering change rates so as to allow the system to create a supersolid pattern even for $a_s\approx a_s^{c}$, which may not be feasible due to i) experimental impossibility of a slow enough ramp speed, or ii) the length such a ramp would require to be longer than the lifespan of the condensate. 

An alternative way, as we propose here, is to create a close-to-transition supersolid by the Bang-Bang shortcut (see Sec. \ref{sec:BangBang}), and to afterward carry it towards the desired supersolid state with a finite ramp velocity. The name ``Parachute Jump'' makes reference to the initial Bang-Bang shortcut, the ``jump'', and the slow $a_s$ reduction afterwards, the ``parachute''.

The initial supersolid formation carries an a priori unnecessary excitation (see Eq. \ref{eq:excite}). However, after the end of the Bang-Bang protocol, the translational symmetry of the superfluid is already broken and the supersolid can continuously morph decreasing the energy through the ramp. Thus, the scheme works by compensating the excitation created by the Bang-Bang with a {subsequent} ``adiabatic'' ramp.

\begin{figure}[t]
 \centerline{\includegraphics[width=1.0\linewidth]{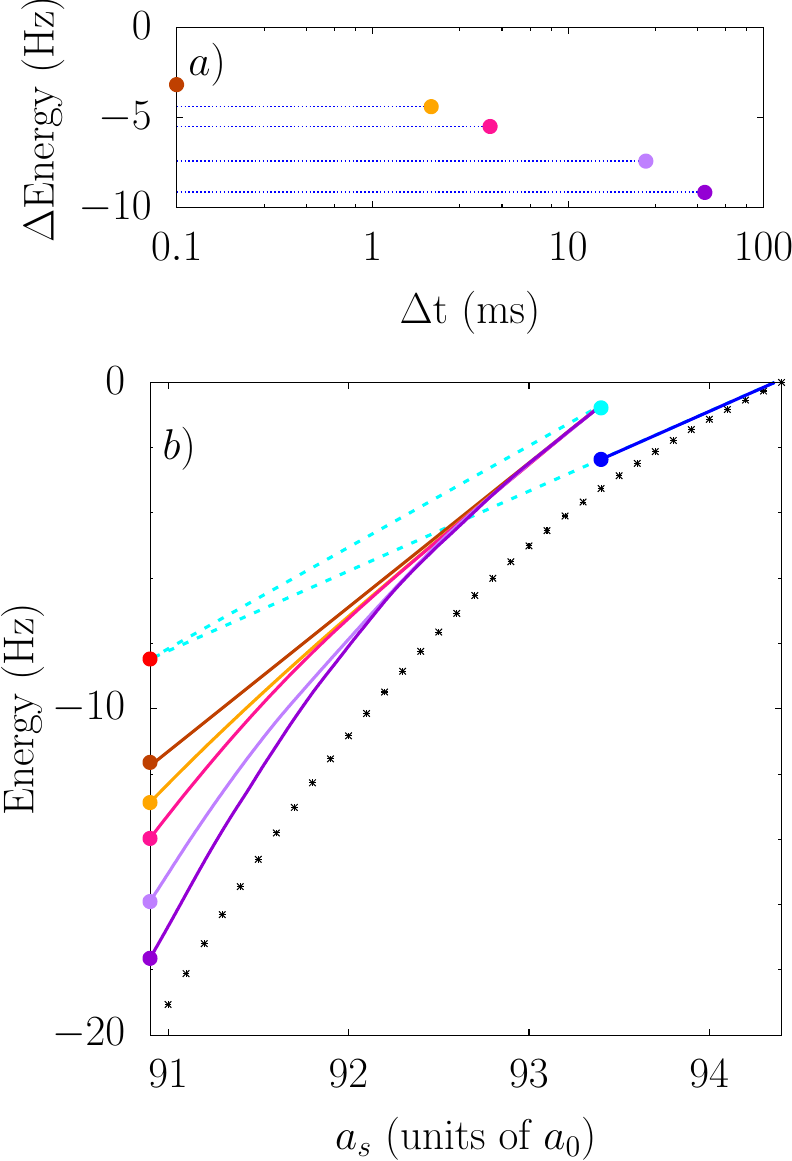}}
 \caption{{The Parachute Jump procedure for different speeds in the last ramp, namely, $v=\infty$ or quench (dark-orange), $v=1$ $a_0/ms$ (orange), $v=0.5$ $a_0/ms$ (magenta), $v=0.1$ $a_0/ms$ (purple), $v=0.05$ $a_0/ms$ (dark-violet). $a$) the energy difference respect to a single quench ($\Delta$Energy) versus the time spent in the last ramp ($\Delta$t) in logarithmic scale. $b)$ Energy plot for the initial Bang-Bang (color scheme of Figs. \ref{fig:IPR},\ref{fig:Energy}) and different ramps. The Bang-Bang applied in all cases is the same, shown also in Figs. \ref{fig:IPR},\ref{fig:Energy}, and the ramps are initiated 3 ms after the second quench.}}
 \label{fig:PJ}
\end{figure}

Fig. \ref{fig:Energy} shows the behavior of the energy along the protocol. The cyan dashed-dotted line is the first part, the Bang-Bang, and the magenta solid one the following ``parachute''. It is apparent that for $da_s/dt=0.04 \ a_0/$ms the final energy is reduced below the single-quench approach. The chosen speed is just an example to show the ideal behavior of the supersolid under the scheme, while for each experimental application, one should find the optimal speed (naturally, slower changes allow the system to adapt better). {Find in Fig. \ref{fig:PJ} examples of other ramp speeds.}

{In experimental realizations of BECs dissipative mechanisms may exist which reduce the energy of the system over time. In such cases, the excitation level of a condensate during a ramp would be dependent both on the speed of the ramp (how adiabatic it is) and on the strength of the dissipative phenomena. If one ignores the dissipation when performing numerical simulations ---as it may be done e.g. to have a clearer picture of the underlaying mechanisms or for a lack of knowledge of the dissipative mechanism of the system--- the ramp would need to excite the system less and be more adiabatic (slow) in order to render a similar excitation level as the experiment. For instance, in \cite{biagioni2022} they found that for their specific realization (which is the one featured in this paper) the measured excitation levels corresponded well with the simulations of slower ramps which were done disregarding dissipative phenomena. Using slower ramps to simulate dissipation may be more problematic if the ramp crosses a phase transition \footnote{{In their case, although they crossed a transition both the experimental and theoretical ramps were shorter than the formation time of the supersolid and thus they did not see an issue due to crossing.}}.}

\begin{figure*}[t!]
    \centerline{\includegraphics[width=2\columnwidth]{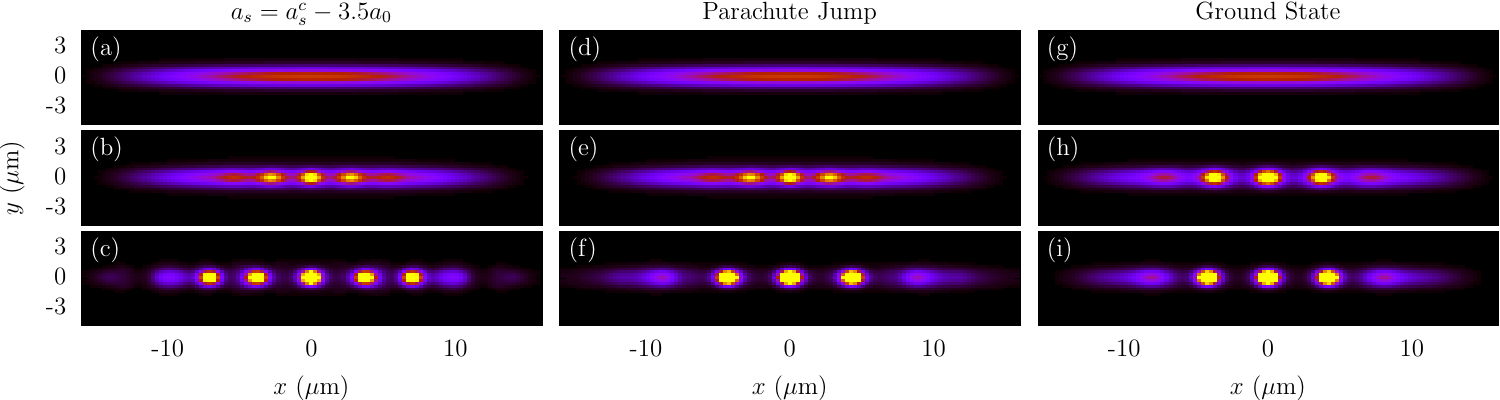}}
\caption{Comparison of the density configurations when applying a single quench (left), the Parachute Jump protocol (center), and the Ground State configuration for $a_s$ of the corresponding Parachute Jump image (right). (a-c) and (d-e) show the density distribution after the quench at $\tau=0$, $\tau=20$ ms, and $\tau=85$ ms of the single quench and the Parachute Jump respectively. (g-i) show the ground state density distribution at $a_s=a_s^c+1.5 \ a_0$, $a_s=a_s^c-1.0 \ a_0$ and $a_s=a_s^c-3.5 \ a_0$, which correspond to the $a_s$ values the system has in panels d, e, and f. The color distribution is the {one used in Fig.\ref{fig:dens_shortcut}} for all panels, for a better comparison.}
\label{fig:dens_parachute}
\end{figure*}

In Fig. \ref{fig:dens_parachute} the {density distributions along} Parachute Jump is compared both to the single quench approach and the ground state. By comparing the initial phase of the supersolid formation ---panels (b) and (e)--- with the corresponding ground state in panel (h) it is apparent that the distance between the maxima is longer for the ground state, i.e., the characteristic momentum of the supersolid is lower than the $k_{\text{rot}}$ of the superfluid. The maxima both in the single quench and the Parachute Jump protocol tend to separate in the evolution, as indicated by panels (c) and (f).

The behavior of the IPR along the scheme (see magenta double dotted dashed line in Fig. \ref{fig:IPR}) shows that not only the collective oscillations are less prevalent than with a single quench ---the IPR oscillations have a significantly smaller amplitude--- but also that the finale IPR value is really close to the one of the ground state contrary to what happens after a single quench. Fig\ref{fig:dens_parachute}(c), (f) and (i) also confirm the higher resemblance to the ground state.

{Regarding the design of this scheme for a particular case one must optimize the first step, the Bang-Bang (see discussion at the end of Sec.\ref{sec:BangBang}). In addition, there is a tradeoff with the speed of the linear ramp reducing the scattering length. Slower ramps would render better energy reduction (the aim of the proposal), but they come at the expense of higher atom loss. Since one could also see an energy reduction when instead of a linear ramp a quench is used, one has the capability to choose the optimal speed for the given experiment. Find in Fig. \ref{fig:PJ} simulation result for Parachute Jumps using different speeds. It is also convenient to keep in mind that $a_s^{\text{obj}}<a_s^c$ must be chosen far from the desired final scattering value since the energy reduction is greater when $a_s^{\text{obj}}$ is further from the final value (see lines in Fig. \ref{fig:PJ}).}

\section{Kick-based pre-solidification (Kick-Bang approach)}\label{sec:KickBang}

The philosophy behind the Bang-Bang method is to increase the spontaneous formation rate of the roton by creating favorable conditions {(see Sec. \ref{sec:BangBang})}. An alternative path is to induce an excitation of the system before crossing the transition. The idea is to mechanically ``kick'' the system so as to excite it in such a way that the roton gets populated and quench it afterward, thus starting the formation process with an already well populated roton mode. Within such framework the system does not strictly experience a phase transition with its spontaneous symmetry breaking, as the symmetry breaking may occur ``artificially'' in the superfluid phase in the form of a mechanical excitation. Ideally, such excited state is later stabilized by reducing the scattering length. 
This process is presented as a proof of principle, with the view to study the characteristics of mechanical excitation and how they may affect the transition between a superfluid and a supersolid. This aim guided the choice of the ``kick'' presented in this manuscript, which instead of targeting the roton specifically has a more general shape.

Choosing the initial superfluid state to be close enough to the transition ---it can always be quenched close to the transition point without almost any excitation--- the excitation spectrum will have a local minimum around $k=k_{\text{rot}}$. In such a case there is no need to even target the roton specifically, as more general kicks will also be able to excite the roton significantly (although specially targeting it will obviously yield higher roton populations). 

In the example of this mechanism presented here, the system is excited by applying a force of $f/k_B\approx 6$nK$/\mu$m (where $k_B$ is the Boltzamann constant) towards positive $y$ in some range of $x$ values, between $8\mu$m and $10\mu$m, and towards negative $y$ between $-8\mu$m and $-10\mu$m \footnote{The inverse direction of the applied forces is done to keep the center of mass stable during the evolution, it is not, however, necessary to see the formation acceleration}. General as it could be, the applied kick must in some way break the translational symmetry along the supersolid formation direction. Otherwise, only the dipole or scissor modes could be excited. 

The force is applied for 2 ms ---enough to deform the condensed gas--- and afterward switched off. The system is left to evolve for 18 ms and then quenched to the ``slow'' formation value of $a_s^c-1.0 \ a_0$. If instead of 18 ms the system evolves less time before the quench the formation shortening may very well occur. However, it has been left to evolve to allow the deformation to spread and thus reduce the dependence on the specific characteristics of the applied kick in the data (see Fig. \ref{fig:dens_shortcut}(g)).

The process creates a significant excitation (see Fig. \ref{fig:Energy}). There is a first excitation period due to the applied kick and a consecutive energy reduction due to the quench. One could in principle engineer the kick to increase the IPR at the moment of the quench, thus reaching a more efficient energy reduction. 

The time elapsed after the kick increases the population of the roton at the moment of the quench with respect to the case of a not kicked condensate. Such an increase in population is, in fact, accelerating the formation process (see  Fig. \ref{fig:IPR}). This can be seen in the value of IPR, which for the Kick-Bang scheme reaches the equilibrium value after 20 ms of being at $a_s<a_s^c$, instead of almost 50 ms it would have required if it was just transferred to $a_s^{\text{obj}}$ without a previous kick. Note that although it oscillates around the equilibrium IPR value as the single quench and the Bang-Bang do, they do not reach the same state, as all of them are different excited states of the supersolid; if dissipation mechanisms were considered they could all converge to the same state after some relaxation time.

Fig. \ref{fig:dens_shortcut} enables us to make a comparison both between this ``scheme'' and the cases of a single quench to $a_s^{\text{obj}}$ and with the Bang-Bang scheme. The modulation apparent in the density distribution coincides with the IPR values of Fig. \ref{fig:IPR}, confirming the acceleration of the formation. 

Note that the configuration of the condensate after the Kick-Bang scheme is slightly different, as it shows a minimum in the center instead of a maximum. This is a consequence of two almost degenerate states being realizable within these parameters. The phase of the density, i.e. where the maximum density points are located, is not relevant for infinite systems, however, when we impose a trap only two configurations survive, the ones seen in Fig. \ref{fig:dens_shortcut}. Small variations in the initial conditions of the superfluid make it select one or the other supersolid configuration.

There is an infinite amount of ``kicking'' protocols that could be implemented, one will expect that one of the most efficient kicks to accelerate the formation process would be to switch on a cosine-like potential with $\lambda=2\pi/k_{\text{rot}}$ for some time (which would directly target the excitation of the roton). {Performed simulations indicate that indeed such kick protocol, which contrary to the presented one does not require to wait after the kick in order to allow the excitations to populate the roton, is efficient in speeding up the evolution with less excitation than a more general kick. Since the aim of this section is to show a proof of principle of how noise can affect the formation process we did not comment the sinusoidal kick, despite its effectiveness.}

\section{Three-body losses}
\label{sec:3B}

{Together with the enabling of the exploration of close-to-transition supersolids, one of the principal advantages of the presented Bang-Bang is the reduction of atom losses and required waiting time during the supersolid preparation procedure. This section devotes itself to analyzing the viability of the Bang-Bang procedure to obtain this loss reduction. Since the aim of Parachute Jump is to reduce the excitation the values of atom loss are not especially relevant ---although they will be commented on at the end of the section--- as neither are they for the Kick-Bang, which is a proof of concept rather than a proposal with an experimental application in mind.}
%

{The atom loss may occur due to various phenomena (e.g. finite size of the trapping potential allowing atoms to escape). In the experiment from which this proposals are inspired they had three-body collisions as the main atom loss mechanism \cite{Ta19}, which are not considered in the derivation of GP equation. Such atom loss can be included in the GP equation by introducing a new term into the Hamiltonian,}
\begin{equation}
    {i\hbar\partial_t\psi=\left[ \hat{H}_{GP}-i\hbar\frac{L_3 N^2}{2}|\psi|^4\right]\psi,}
\end{equation}
{where $L_3$ is a positive constant which we call ``loss rate'' and $\hat{H}_{\text{GP}}$ is the Hamiltonian given by the functional differentiation of the energy functional [see Eq.\eqref{eq:GPenergy}].}

{Calculating the loss rate is a challenging task since it relies on the bound state of two particles that are formed during a collision. Although some predictions explicitly dependent on $a_s$ have been made (see e.g.\cite{bedaque2000three,d2005scattering}), there is no general formula to obtain the loss rates, and must thus be evaluated (be it theoretically or experimentally) for each particular case of interest but for extreme cases such as very high $a_s$.} 

{The schemes presented here rely on manipulating the roton instability and can be applied to various physical setups, regardless of their atom loss rates, as long as they exhibit a roton instability that drives a phase transition. However, it is useful to examine the example of atom loss in the system of $^{162}$Dy atoms that was presented.}

{Although the scaling of the loss rate has already been investigated in the case of dipolar particles \cite{ticknor2010three}, the specific region around the superfluid-supersolid phase transition, where characteristic scattering lengths of contact and dipolar interactions are similar, is difficult to compute. In 2019 Tanzi \textit{et al.} noted that for the specific case of $^{162}$Dy the loss rate was not known, and thus they performed experimental measures of it in various regimes \cite{Ta19}, providing the value $L_3\approx 2\times10^{-28}$ cm$^6/$s \footnote{{They performed two sets of measures rendering the values $2.5\times10^{-28}$ cm$^6/$s and $2.1\times10^{-28}$ cm$^6/$s.}}. Since posterior experiments with this isotope around the relevant scattering values for the superfluid-supersolid phase transition did not observe any anomalous behavior regarding the atom loss rate when changing $a_s$, there is nothing indicating that the quenches and linear ramps used in the proposals would affect the $L_3$ significantly.}

{The total atom number change rate can be written as,}
\begin{equation}
    {\dot{N}/N=-L_3\langle n^2\rangle=-L_3N^2\langle|\psi|^4\rangle=-L_3N^2\int|\psi|^6dV,}
\end{equation}
{thus, the atom loss is directly related to the integral of $|\psi|^6$\footnote{{As a matter of fact, this value may also be referred to as an IPR \cite{cuevas2002critical}. We abstain from doing so in order to avoid confusion.}}. During the formation process, $\langle |\psi|^4\rangle$ of the condensate (which has similar behavior to IPR in Fig. \ref{fig:IPR}) is close to the initial one, with some small oscillations, which because their small module are not relevant for the total atom loss. The performed simulations predict $\langle |\psi|^4\rangle\approx 1.5\times10^{-5}$ 1/$\mu$m$^6$ for the superfluid case. Although atom losses could modify $\langle |\psi|^4\rangle$, the steady state can be changed by the number of atoms after all \cite{eberlein2005exact}, their effect would be low according to GP-simulations. Then $\langle |\psi|^4\rangle$ can be approximated as stable during the formation process regarding the effect on atom loss, rendering,}
\begin{equation}
    {N(t)=N_0/\sqrt{1+2N_0^2L_3 \langle |\psi|^4\rangle t},}
\end{equation}
{where $N_0$ is the initial number of atoms. Plugging in the relevant numbers one sees that the atom losses in the single quench and Bang-Bang schemes are,}
\begin{equation}
    {\Delta N_{\text{Quench}}\approx 3400, \quad \Delta N_{\text{Bang-Bang}}\approx 1500.}
\end{equation}
{Therefore, proving the positive effect of the Bang-Bang scheme. When the system forms the supersolid $\langle|\psi|^6\rangle$ gets bigger, which accelerates atom loss.}

{The procedure could fail to obtain a supersolid if the number of atoms at the end is such that the roton is stabilized [see Eq.\eqref{eq:dispersion}]. However, in that scenario the single quench would neither obtain a supersolid, as losses are higher for it. Furthermore, for some cases the Bang-Bang would generate supersolids where the single quench was not able to.}

{Regarding the Parachute Jump, the same applies: The main issue with the atom number is that at the end of the procedure, there should be enough atoms so that the supersolid remains energetically favorable. The atom loss of the ramp must then also be included into account. Fortunately it is not necessary to have an adiabatic ramp to obtain an energy reduction and thus the speed can be tuned to accommodate the atom preservation constraint.}

{It is worth remarking that the proposed protocols are quite general and could be applied to cases in which the particle loss has a significantly different $L_3$ or where the main loss mechanism is not three-body losses. Thus, the discussed numbers for particle losses may change for different systems.}

\section{Conclusions}
\label{sec:conclusions}

As shown, it is possible to reduce the formation time of a supersolid by changing the scattering length to lower values before arriving at the desired value. The losses in the particle number that are present along the formation process can be thus reduced, extending the time left in which to experiment with the supersolid. {The atom loss has been estimated for the specific example of $^{162}$Dy isotope, although noting the generality of the proposals.} It also extends the range of supersolids that are experimentally viable, namely the ones really close to the transition point which would require too long times to form. 

When the desired supersolid is well below the transition point, the excitation may be huge if the transition is not driven in a slow enough manner. One should ideally drive the system as slowly as needed to allow the density to redistribute, adiabatically. This, however, is hardly possible due to the slow speeds it would require close to the transition point, where the formation time gets large \cite{Ch18,alana2022crossing,alana2023formation}. The slow enough speeds could be experimentally inaccessible, or the required time so long that the three-body losses would eliminate the condensate altogether. Precisely to address this issue has the Parachute Jump scheme been proposed. 

Our proposal is to use the Bang-Bang approach to accelerate the formation of the supersolid close to the transition and then reduce the scattering length by a ``slow'' ramp. Once the supersolid pattern has formed the system is able to adjust to new scattering values swiftly, which allows a quasi-adiabatic driving towards the endpoint. {Nevertheless, following Eq.\eqref{eq:excite} we noted that fast ramps and quenches may also reduce the overall energy, thus allowing us to choose the Parachute Jump's speed regarding the specific characteristics of the experiment, slowing the ramp if atom loss is low and speeding it up if atom loss is high, since for this scheme there is a tradeoff between excitation and atom loss.}

Both the Bang-Bang and the Parachute Jump schemes represent the most simple, yet effective, optimization techniques for the superfluid-supersolid phase transition. {Within the limitation of both proposals optimization has been discussed using the quasi-1D infinite model \cite{alana2023formation,blakie2020variational}, noting also the range of applicability both of the model and the scheme.} One can of course follow the philosophy of these proposals with more complex schemes. Another way forward in the path of transition optimization could be to combine the dynamical change of scattering length with another dynamical change of the trapping potential. 

We also investigate how a mechanical kick before the transition affects the formation time. We find that even with a kick not aimed to excite any particular momentum state the roton momentum gets populated, which accelerates the formation of the supersolid pattern. This indicates that in addition to the nonzero temperatures present in the experiments, which would be linked to a greater roton population \cite{sanchez2023heating}, mechanical noise may also have a role in explaining the shorter formation times with respect to the mean-field calculations. 

Moreover, the outcomes of the Kick-Bang scheme provide a pathway to assess the impact of mechanical noise on a dipolar gas, which is challenging to estimate through theoretical means. To achieve this, one can compare the time it takes for a supersolid to form under specific scattering conditions after a sudden change, focusing on two scenarios: i) when the initial superfluid is near the transition point (displaying a distinct local minimum around the roton), and ii) when the initial superfluid possesses a scattering length significantly above the transition point and lacks a distinct local minimum apart from $k=0$. If there is a significant difference in the formation times between these cases, it may suggest that the roton mode is being influenced by mechanical noise, providing an additional check of the accuracy of the experiment.

Although the schemes have been presented only in the framework of the superfluid to supersolid phase transition, the philosophy behind them is more general and they could easily be adapted for other phase transitions driven by modulation instabilities in which the researcher, by means of experimental parameters, is able to tune one of the following: 1) the initial population of the unstable modes, 2) the imaginary part of the unstable frequency. An example outside of the dipolar atoms in which these schemes could be implemented may be the soliton creation after an interaction quench in a two-component Bose condensate (see e.g. \cite{cidrim2021soliton}).

\

\begin{acknowledgments}
This work was supported by Grant PID2021-126273NB-I00 funded by MCIN/AEI/ 10.13039/501100011033 and by ``ERDF A way of making Europe'', by the Basque Government through Grant No. IT1470-22, and by the European Research Council through the Advanced Grant ``Supersolids'' (No. 101055319). I thank Prof. Giovanni Modugno, Luca Tanzi, Nicolò Antolini and Giulio Biagioni alongside all people from the Dy-lab at Pisa for interesting discussions, as well as Luca Cavicchioli from the K-Rb lab at Firenze. I also thank my Ph.D. supervisors Michele Modugno and Iñigo L. Egusquiza from the University of the Basque Country for useful feedback.
\end{acknowledgments}

\end{document}